\documentclass[12pt,final]{amsart}
\usepackage{amsmath,amssymb}

\usepackage{color}
\usepackage{soul,graphicx}

\usepackage[color]{showkeys}
\definecolor{refkey}{rgb}{0,0,1}
\definecolor{labelkey}{rgb}{1,0,0}

\newcommand{\calS}{\mathcal{S}}

\newcommand{\calP}{\mathcal{P}}

\newcommand{\calO}{\mathcal{O}}

\newcommand{\dt}{\delta}

\newcommand{\gm}{\gamma}

\newcommand{\Gm}{\Gamma}
\newcommand{\Dt}{\Delta}

\newcommand{\diag}{\mathop{\rm diag}}

\newcommand{\tm}{\times}

\newcommand{\sbs}{\subset}

\newcommand{\wdt}{\widetilde}

\newcommand{\iy}{\infty}

\newcommand{\bF}{\mathbb{F}}

\newcommand{\bZ}{\mathbb{Z}}

\newtheorem{theorem}{\bf  Theorem}
\newtheorem{remark}{ \sc Remark}



\begin{document}
	\begin{center}
		{\bf Random Generator of Orthogonal Matrices in Finite Fields}\\[5mm]
		Lasha Ephremidze and Ilya Spitkovsky\\ [2mm]
New York University Abu Dhabi,  UAE;\;\; E-mail: {\em le23@nyu.edu}
	\end{center}

	\vskip+0.6cm

	{\bf Abstract.} We propose a superfast method for constructing orthogonal matrices $M\in\calO(n,q)$ in finite fields $GF(q)$. It can be used to construct $n\tm n$ orthogonal matrices in $Z_p$ with very high values of $n$ and $p$, and also orthogonal matrices with certain circulant structure.
	Equally well one can construct paraunitary filter banks or wavelet matrices over finite fields. 
		The construction mechanism is highly efficient, allowing for the complete screening and selection of an orthogonal matrix that meets specific constraints. For instance, one can generate a complete list of orthogonal matrices with given $n$ and $q=p^m$ provided that the order of $\calO(n,q)$ is not too large. 	
	 Although the method is based on randomness, isolated cases of failure can be identified well in advance of the basic procedure's start. 
		
		The proposed procedures are based on the Janashia-Lagvilava method which was developed for entirely different task, therefore, it may seem somewhat unexpected. 
		
	\vskip+0.6cm
	
	
	\section {Introduction.} 
	
	Spectral factorization is a powerful mathematical tool with various applications in control engineering and communications. The Janashia-Lagvilava method \cite{JL99}, \cite{IEEE2011} has been developed over the years to solve the challenging spectral factorization problem in the matrix case. This method has been successfully algorithmized \cite{IEEE2018} and offers several advantages over existing methods, as demonstrated in previous works \cite{Robinson}, \cite{TRMI22}. One noteworthy advantage of this method, which can be achieved through a suitable modification, is that it unexpectedly has the ability to generate a multitude of orthogonal matrices over finite fields in mere milliseconds, potentially finding applications in coding theory.
	
	It had been revealed in \cite{JLE13}, \cite{EL2014}, that the Janashia-Lagvilava method was closely related with wavelet theory. 
  The discovery of this connection paved the way for the accomplishments detailed in this paper. Furthermore, the method has been recently generalized for any field satisfying the minimal requirements \cite{GMJ22}. However, these requirements necessitate the existence of a positive element in the field, precluding finite fields. In fact, the field of rational polynomials in several complex variables was the main focus of our paper \cite{GMJ22}, and by taking this viewpoint, we were able to generalize the Janashia-Lagvilava method to matrix functions on the multidimensional torus \cite{JMAA22}. The main result of \cite{GMJ22}, Theorem 4.1, does not hold for finite fields and, accordingly, the procedure described therein cannot guarantee the construction of the required unitary matrix polynomial for finite fields. Nevertheless, as it was further realized and confirmed by numerical simulations, the isolated cases of failure do not exclude the existence of  statistically robust method of constructing above mentioned unitary matrix polynomials for finite fields, while these matrix polynomials can be converted into orthogonal matrices. To be specific, this paper provides efficient computational procedures for proving the following 
	
	\begin{theorem}
		Let $\bF$ be a finite field and $G$ be an $n\tm n$  matrix function of the form
	\begin{equation}\label{Ft}
	G(t)=\begin{pmatrix}1&0&0&\cdots&0&0\\
	0&1&0&\cdots&0&0\\
	0&0&1&\cdots&0&0\\
	\vdots&\vdots&\vdots&\vdots&\vdots&\vdots\\
	0&0&0&\cdots&1&0\\
	\zeta_{1}(t)&\zeta_{2}(t)&\zeta_{3}(t)&\cdots&\zeta_{n-1}(t)&1
	\end{pmatrix},
	\end{equation}
where $\zeta_i(t)=\sum_{k=1}^N\gm_{ik}t^{-k}$, $i=1,2,\ldots,n-1$, are Laurent polynomials, $0,1,\gm_{ik}\in\bF$.  Denote by $\Gm_i$, $i=1,2,\ldots,n-1$, the Hankel matrices
\begin{equation}\label{Gm}
\Gamma_i=\begin{pmatrix}0&\gamma_{i1}&\gamma_{i2}&\cdots&\gamma_{i,N-1}&\gamma_{iN}\\
\gamma_{i1}&\gamma_{i2}&\gamma_{i3}&\cdots&\gamma_{iN}&0\\
\gamma_{i2}&\gamma_{i3}&\gamma_{i4}&\cdots&0&0\\
\cdot&\cdot&\cdot&\cdots&\cdot&\cdot\\
\gamma_{iN}&0&0&\cdots&0&0\end{pmatrix},
\end{equation}
 and by $\Dt$  the following $(N+1)\tm (N+1)$ matrix
\begin{equation*}\label{Dt}
\Dt=\sum_{i=1}^{n-1}\Gamma_i^2+I_{N+1}.
\end{equation*}	
If $\det\Dt\not=0$,  then there exists  a unique $n\tm n$ polynomial matrix $U$,
\begin{equation}\label{Ut}
{\bf U}(t)=\begin{pmatrix}u_{11}(t)&u_{12}(t)&\cdots&u_{1,n-1}(t)&u_{1n}(t)\\
u_{21}(t)&u_{22}(t)&\cdots&u_{2,n-1}(t)&u_{2n}(t)\\
\vdots&\vdots&\vdots&\vdots&\vdots\\
{u_{n1}(t)}&{u_{n2}(t)}&\cdots&{u_{n,n-1}(t)}&{u_{nn}(t)}\\
\end{pmatrix},
\end{equation}
$u_{ij}(t)=\sum_{k=0}^N a_{ij,k}t^k$, $a_{ij,k}\in\bF$, such that the entries in the product matrix $$G(t)\diag(1,1,\ldots,1,t^{-N}){\bf U}(t)$$
contain only the non-negative powers of $t$ $($i.e. the coefficients of  $t^{-k}$, $k=1,2,\ldots$, are $0$ of $\bF$\,$)$ and, in addition, $U(1)=I_n$.

Furthermore,  the matrix polynomial \eqref{Ut} has the property that
\begin{equation}\label{UUT1}
{\bf U}(t)\wdt{{\bf U}}(t)=I_n
\end{equation}
where  $\wdt{{\bf U}}(t)=\sum_{k=0}^NU_k^Tt^{-k}$  for ${\bf U}(t)=\sum_{k=0}^NU_kt^k$.
	\end{theorem}
The straightforward computations show that, due to the property \eqref{UUT1},  the $n\tm n$ matrix 
\begin{equation*}\label{W0}
W_0={\bf U}(-1)=U_0-U_1+U_2-\ldots (-1)^NU_N,
\end{equation*}
and the block circulant $n(N+1)\tm n(N+1)$ matrix $W$ with the first (block) row
\begin{equation}\label{W1}
W_1=(U_0\;U_1\;\ldots\;U_N)
\end{equation}
are orthogonal, i.e.
\begin{equation}\label{W0W}
W_0W_0^T=I_n\;\; \text { and }\;\; WW^T=I_{n(N+1)}.
\end{equation}
Hence, relying on the proof of Theorem 1, after selecting positive integers $n$ and $N$, one can randomly generate the coefficients $\gm_{ik}$, $1\leq i<n$, $1\leq k\leq N$, of polynomials $\zeta_i$ in \eqref{Ft} from the field $\bF$ and construct the orthogonal matrices satisfying \eqref{W0W}. Concerning computer memory allocation, these $(n-1)N$ integers, $\gamma_{ik}$ (generators), contain full (lossless) information about $n^2(N+1)$ integers of $W_1$ in \eqref{W1}. Explicit formulas converting the matrix $W_1$ back into the generators are also provided. 

\begin{remark}
	Note that the method generates only the trivial matrix $W_0=I_n$ for $q=2$ since $-1=1$ in $GF(2)$ and ${\bf U}(1)=I_n$ by construction. However, the circulant matrix $W$ still may be different from the trivial matrix $I_{n(N+1)}$.
\end{remark}

We would like to emphasize that Janashia-Lagvilava method in fact contains the proof of the above theorem and it can be readily apparent from previous publications \cite{IEEE2011}, \cite{GMJ22}. However, the primary theoretical improvement of the Janashia-Lagvilava method proposed in what follows is the modification of the existing computational procedures. This modification eliminates the need to invert any $n\tm n $ matrix, which was previously required (see \cite[Eq. (51)]{IEEE2011} or reasoning after Eq. (4.23) in \cite{GMJ22}), and makes the entire procedure superfast. Namely, constructing orthogonal matrices $W_0$ and $W$ of sizes $n\tm n$ and $n(N+1)\tm n(N+1)$, respectively, requires $O(N^3)+(N+1)^2n^2$ (with the exact constant 1 instead of $O$ in front of the second summand)  multiplications (in the field $\bF$\,).  Furthermore, for arbitrarily large $n$, it is possible to select the second parameter $N$ as small as $1$, admitting a moderate level of randomness.

\section{Notation }

This section summarizes the notation used throughout the paper, although most symbols have already been introduced in the introduction. Let $\bF\equiv GF(q)=GF(p^m)$ be the Galois field consisting of $q=p^m$ elements, where $p$ is a prime number, and $\bZ_p\cong\bZ$  $(\mod p)$. Wherever we encounter 0 or 1 they are presumed to belong to this field, while $-1$ is the additive  inverse of $1$ in $\bF$.

 For a set $\calS$, let $\calS^{n\tm k}$ be the set of $n\tm k$ matrices with entries from $\calS$, while $\calS^n:=\calS^{n\tm 1}$. $I_n=\diag(1,1,\ldots,1)\in\bF^{n\tm n}$ is the identity matrix and $0_{n\tm k}\in\bF^{n\tm k}$ is the $n\tm k$ matrix consisting of zeros.  A matrix $M\in\bF^{n\tm k}$ with entries $M_{ij}\in \bF$ is denoted by $M:=[M_{ij}]_{i=1:n}^{j=1:k}$, or simply by $M:=[M_{ij}]$. A matrix $M=[M_{ij}]\in \bF^{n\tm n}$ is called orthogonal if 
 \begin{equation}\label{MMT}
 MM^T=I_n,
 \end{equation}
where $M^T=[M_{ji}]$ stands for the transpose of $M$. The above equation implies also that $M^TM=I_n$.
The set of orthogonal $n\tm n$ matrices in $GF(q)$ is denoted by $\calO(n,q)$.

Let $\calP\equiv\calP[\bF]$ be the set of Laurent polynomials with coefficients from $\bF$:
 \begin{equation}\label{calP}
\calP:=\left\{\sum\nolimits_{k=k_1}^{k_2} c_kt^k: c_k\in\bF,\;k_1,k_2\in\bZ;\;k_1\leq k_2   \right\}.
 \end{equation}
We also consider the following subsets of $\calP$: $\calP^+$, $\calP^-$, $\calP^+_N$ and $\calP^-_N$, where $N$ is a non-negative integer,  which corresponds to sets $k_1=0$, $k_2=0$, $0=k_1\leq k_2=N$, and $-N=k_1\leq k_2=0$ in \eqref{calP}, respectively. So, $\calP^+$ is the set of usual polynomials, and $\calP^+_N$ is the set of polynomials of degree less than or equal to $N$.  Of course, any polynomial $P\in\calP[\bF]$,
 \begin{equation}\label{pt}
P(t)=\sum\nolimits_{k=k_1}^{k_2} c_kt^k,
\end{equation} 
  can be considered as the function $P:\bF\to\bF$ by plugging in it the values from $\bF$ instead of indeterminate variable $t$. The coefficient $c_0\in\bF$ is called the free term of $P$.  It is also naturally assumed that $\bF\sbs\calP[\bF]$, i.e. $c\in \bF$ can be identified with the  (constant) polynomial $ct^0$. Saying that a polynomial \eqref{pt} is constant means that all coefficients, except for a possible term $c_0$, are zero. In this case, slightly abusing notation, we can write $P(t)=c_0$. In a similar manner, $P\in\calP^+$ means that all the coefficients of $P$ corresponding to  negative powers of $t$ are equal to $0$.
  
A matrix (Laurent) polynomial ${\bf P}\in \calP^{n\tm k}$ can also  be viewed as a polynomial with matrix coefficients,
 \begin{equation}\label{Pt}
{\bf P}(t)=\sum\nolimits_{k=k_1}^{k_2} C_kt^k,\;\;\; C_k\in\bF^{n\tm k}.
\end{equation}

For a polynomial \eqref{pt}, let
 \begin{equation}\label{wdtp}
  \wdt{{\bf P}}(t)=\sum\nolimits_{k=k_1}^{k_2} c_kt^{-k},
\end{equation} 
and for a matrix polynomial \eqref{Pt}, let 
 \begin{equation}\label{wdtP}
\wdt{{\bf P}}(t)=\sum\nolimits_{k=k_1}^{k_2} C_k^Tt^{-k},
\end{equation}
i.e. if ${\bf P}(t)=[{\bf P}_{ij}(t)]_{i=1:n}^{j=1:k}$, then $\wdt{{\bf P}}(t)=[\wdt{{\bf P}}_{ji}(t)]_{j=1:k}^{i=1:n}$. A matrix polynomial ${\bf A}$ is called {\rm paraunitary} if 
\begin{equation}\label{AAT}
{\bf A}(t)\wdt{{\bf A}}(t)=I_n,
\end{equation}
i.e., all matrix coefficients of ${\bf A}\wdt{{\bf A}}$ are $0_{n\tm n} $ except the free term, and the latter is equal to $I_n$. Thus, the matrix polynomial ${\bf U}$ in \eqref{UUT1} is paraunitary.

The symbol $\dt_{ij}$ stands for the Kronecker delta, i.e., $\dt_{ij}=1$ if $i=j$ and $\dt_{ij}=0$ otherwise. 

Note that basic linear algebra rules apply to linear spaces over finite fields, and therefore we can use them without further justification.

\section{Constructive proof of Theorem 1 }

We mainly follow the path developed in the Janashia-Lagvilava method described in publications \cite{IEEE2011}, \cite{GMJ22}. Therefore, we do not repeat many details given therein and emphasize only the arising differences. 

Construction of the paraunitary matrix \eqref{Ut}, for a given matrix \eqref{Ft},  starts by considering the following system:
\begin{equation}\label{S1}
\begin{cases} \zeta_1x_n-\wdt{x_1}\in \mathcal{P}^+,\\
\zeta_2x_n-\wdt{x_2}\in \mathcal{P}^+,\\
\vdots\\
\zeta_{n-1}x_n-\wdt{x_{n-1}}\in \mathcal{P}^+,\\
\zeta_1x_1+\zeta_2x_2+\ldots+\zeta_{n-1}x_{n-1}
+\wdt{x_n}\in \mathcal{P}^+.
\end{cases}
\end{equation}
We search for its solution in the class $\calP^+_N$, i.e., for $(x_1,x_2,\ldots,x_n)\in (\calP^+_N)^{1\tm n}$ satisfying the conditions of the system \eqref{S1}.

It is proved in  \cite{IEEE2011}, \cite{GMJ22} that if we find $n$ different solutions 
\begin{equation}\label{Xj}
{\bf U}_j(t)=(x^j_1(t), x^j_2(t),\ldots,x^j_n(t))^T\in (\calP^+_N)^{n\tm 1}, \; j=1,2,\ldots,n,
\end{equation}
of \eqref{S1} satisfying also 
\begin{equation}\label{X1}
{\bf U}_j(1)=(\dt_{1j}, \dt_{2j}, \ldots,\dt_{nj})^T,
\end{equation}
 and  assume
\begin{equation*}	
\hat{{\bf U}}_j(t)=(x^j_1(t), x^j_2(t), \ldots, x^j_{n-1}(t),\; t^N\wdt{x^j_n}(t))^T, \;\; j=1,2,\ldots,n
\end{equation*}
(the order of coefficients has been reversed in the last entry of ${\bf U}_j$  in order to get  $\hat{{\bf U}}_j$), then \eqref{Ut} can be represented as a concatenation of the columns 
\begin{equation}\label{U2}	
{\bf U}=(\hat{{\bf U}}_1, \hat{{\bf U}}_2, \ldots,   \hat{{\bf U}}_n),
\end{equation}
and thus constructed matrix \eqref{U2} will satisfy the condition \eqref{UUT1}   as well.

To solve the system \eqref{S1}, we rewrite it in the equivalent form in terms of the coefficients of polynomial entries. In particular, to obtain the solution \eqref{Xj} satisfying  \eqref{X1}, we reduce \eqref{S1} to 
\begin{equation}\label{S2}
\begin{cases}
D\cdot{X_1}-\Gamma_1\cdot X_n={\bf 0}, \\
D\cdot{X_2}-\Gamma_2\cdot X_n={\bf 0}, \\
\cdot\;\;\;\;\;   \cdot\;\;\;\;\;    \cdot\\
D\cdot{X_j}-\Gamma_j\cdot X_n={\bf 1}, \\
\cdot\;\;\;\;\;   \cdot\;\;\;\;\;    \cdot\\
D\cdot{X_{n-1}}-\Gamma_{n-1}\cdot X_n={\bf 0}, \\
\Gamma_1\cdot X_1+\Gamma_2\cdot X_2+\ldots+\Gamma_{n-1}\cdot
X_{n-1}+D\cdot{X_n}={\bf 0}, \end{cases}
\end{equation}
$j=1,2,\ldots, n$, where
$$
D=\begin{pmatrix}1&1&1&\cdots&1&1\\
0&1&0&\cdots&0&0\\
0&0&1&\cdots&0&0\\
\cdot&\cdot&\cdot&\cdots&\cdot&\cdot\\
0&0&0&\cdots&0&1\end{pmatrix},\;\;
\Gamma_i=\begin{pmatrix}0&0&0&\cdots&0&0\\
\gamma_{i1}&\gamma_{i2}&\gamma_{i3}&\cdots&\gamma_{iN}&0\\
\gamma_{i2}&\gamma_{i3}&\gamma_{i4}&\cdots&0&0\\
\cdot&\cdot&\cdot&\cdots&\cdot&\cdot\\
\gamma_{iN}&0&0&\cdots&0&0\end{pmatrix},
$$
${\bf 0}=0_{(N+1)\tm 1}=(0,0,\ldots,0)^T$, ${\bf 1}=(1,0,0,\ldots,0)^T\in\bF^{N+1}$ are given coefficients, and
$$
X_i=(a_{i0},a_{i1},\ldots,a_{iN})^T, \text{ where } x_i(t)=\sum\nolimits_{k=0}^Na_{ik}t^k,
$$
are unknown matrix coefficients. Note that matrices $D$ and $\Gm_i$ differ from the corresponding matrices \cite[Eq. (26)]{IEEE2011}, \cite[Eq. (4.10)]{GMJ22}  in the first row. This replacement is justified by freedom of selection of constant terms of polynomials entering \eqref{S1}. On the other hand, this flexibility allows the condition ${\bf U}(1)=I_n$ to be merged into the system \eqref{S1}. Therefore, we can avoid an expensive step of inverting $n\tm n$ matrix required in preceding papers.  Note also that $\Gm_i$ differs from \eqref{Gm} in the same way, however, this difference is immaterial and we keep the notation for convinience.

We proceed solving the system \eqref{S2} in a natural way: we have
\begin{equation}\label{Xi}
X_i=D^{-1}\cdot \Gm_i\cdot X_n+\dt_{ij}D^{-1}\cdot{\bf 1}, \;\;i=1,2,\ldots, n-1.
\end{equation}
Substituting these equations into the last equation of \eqref{S2}, we get
$$
\Gamma_1 D^{-1}\Gm_{1}X_n+\Gamma_2 D^{-1}\Gm_{2}X_n+\ldots+\Gamma_{n-1}
D^{-1}\Gm_{n-1}X_n+D{X_n}=-\Gm_jD^{-1}{\bf 1}
$$
(where it is assumed that the right-hand side is equal to ${\bf 1}$ for $j=n$), which leads to 
\begin{equation}\label{DtX}
\Dt X_n= -D^{-1}\Gm_jD^{-1}{\bf 1},
\end{equation}
where 
$$
\Dt=\sum\nolimits_{i=1}^{n-1}(D^{-1}\Gm_i)^2+I_{N+1}.
$$
The sufficient condition, $\det\Dt\not=0$, for the existence of the solution to \eqref{DtX} is reflected (in the slightly different but equivalent form) in the hypothesis of the theorem. Therefore, assuming that this condition holds, we determine $X_n$ from \eqref{DtX} and then $X_i$, $i=1,2,\ldots,n-1$ from \eqref{Xi}.

\section{Paraunitary filter banks and wavelets over $\bF$}

The theory of paraunitary filter banks (PFB) in real and complex fields was developed in \cite{Vai} and it is closely related with the concept of wavelet matrices as it is presented in \cite{RW}. Indubitably, this theory has many applications in signal processing. A complete parameterization of such filter banks and wavelet matrices depending on Janashia-Lagvilava method was proposed in \cite{EL2014}.

In \cite{Vai2}, PFB were considered over finite fields. It was emphasized that filter banks over finite fields have the advantage that all the round-off error and the coefficient quantization error can be eliminated completely. In addition, its potential applicability in cryptography, in the theory of error-correcting codes, was mentioned. Accordingly, various theoretical properties of PFB were explored. In parallel, the theory of wavelet transforms over finite fields was proposed by several authors \cite{Vet1}, \cite{Cooklev2}, \cite{Poor}, \cite{Fekri} and further applications envisioned. As one can observe, these two developments mostly differ from each other  by terminology only. 

As paraunitary filter banks are nothing but the coefficients of polynomial matrices \eqref{Ut} with property \eqref{UUT1}, the present paper offers a superfast method for constructing PFB and related wavelet matrices. Furthermore, a great deal of facts observed in \cite{EL2014}, \cite{GMJ22} can be directly extended to finite fields setting as well. In particular, whenever paraunitary matrix 
\begin{equation}\label{Ut2}
{\bf U}(t)=\sum\nolimits_{k=0}^NU_kt^k
\end{equation}
is constructible by the method described in the paper, it is unique for a given set of generators, its determinant is always equal to $t^N$, i.e.
\begin{equation}\label{det1N}
\det {\bf U}(t)=t^N,
\end{equation}
and the last row of the matrix $U_N$ in \eqref{Ut2} cannot be all zeros. Conversely,  if we have a paraunitary matrix \eqref{Ut2} such that $(n,j)$-th entry of $U_N$ differs from zero, $(U_N)_{nj}\not=0$, then the generators $\zeta_i$, $i=1,2,\ldots,n-1$, can be reconstructed by the formulas (cf. \cite[Eq. (25)]{EL2014})
\begin{equation}\label{zeta}
\zeta_i(t)=\left[\wdt{u_{ij}}(t)\big(t^N \wdt{u_{nj}}(t) \big)^{-1}\right]^-, \;\;\;i=1,2,\ldots,n-1,
\end{equation}
where under $\big(t^N \wdt{u_{nj}}(t) \big)^{-1}$ we assume a formal inverse series of the polynomial $t^N \wdt{u_{nj}}(t)$ (it exists since its constant term differs from 0), and $\left[\sum_{k=-N}^\iy a_kt^k\right]^-:=\sum_{k=-N}^{-1} a_kt^k$. The formula \eqref{zeta} can be derived similarly to \cite[Eq. 56]{EL2014}.

\section{Numerical simulations}

We performed numerical simulations described in this section using a MATLAB code that was executed on a laptop equipped with an Intel(R) Core(TM) i7 8650U CPU (with a clock speed of 1.90 GHz) and 16.00 GB of RAM.  Whenever the matrix parameters exceeded the available memory of our local computer, we utilized a compute node from the High Performance Computing resources at New York University, Abu Dhabi. The compute node had an Intel(R) Xeon(R) CPU E5-2680 v4 @ 2.40GHz processor, 28 CPUs, and 102 Gigabytes of usable memory. The results of these simulations are given in Tables 1 and 2, respectively.  
They present the computational time ($t$ in seconds) required to construct an $n\tm n$ orthogonal matrix $W_0$ or an $n(N+1)\tm n(N+1)$ circulant orthogonal matrix $W$ (see \eqref{W0W}) with elements from $Z_p$ by the proposed method.  The corresponding columns of the tables display the values of $p$, $n$, $N$, and $t$. (The gaps in the last row of Table 2 indicate that the available RAM was not sufficient to perform the corresponding computations.)
 We also statistically estimate the probabilities of failure for the construction of the corresponding  matrices. Namely, for each triple $(p,n,N)$ we conducted  1000 random trials and recorded the number of failures $f$. The observed stability of results, as shown in the $f$ column of the table, indicates that the probability of failure is contingent solely upon $p$. 
At first glance, this appears particularly surprising, warranting further investigation into its underlying cause.

\newpage

\begin{center}
	{\footnotesize Table 1 \\Simulation Results for Orthogonal Matrix Generation  on Local Laptop \par}
\end{center}
{\fontsize{8}{8pt}\selectfont
	\begin{center}
		\begin{tabular}{|c|c||c|c|c||c|c|c||c|c|c||c|c|c|}
			\hline &&&&&&&&&&&&&\\[-2mm]
	$n$&$N$&$p$&$t$&f &$p$&$t$&f &$p$&$t$&f &$p$&$t$&f \\
			\hline &&&&&&&&&&&&&\\[-2mm]
	100&1&7&0.04&141&97&0.04&14&997&0.07&1&4999&0.81&0\\
			\hline &&&&&&&&&&&&&\\[-2mm]
	100&10&7&0.06&132&97&0.06&7&997&0.23&1&4999&4.59&0\\
		    \hline &&&&&&&&&&&&&\\[-2mm]
	100&50&7&0.23&144&97&0.28&17&997&1.14&0&4999&17.1&0\\
			\hline &&&&&&&&&&&&&\\[-2mm]
	500&1&7&0.07&129&97&0.07&12&997&0.10&1&4999&0.87&1\\
			\hline &&&&&&&&&&&&&\\[-2mm]
	500&10&7&0.15&145&97&0.18&10&997&0.37&2&4999&4.72&0\\
			\hline &&&&&&&&&&&&&\\[-2mm]
	500&50&7&0.83&155&97&0.98&12&997&1.92&3&4999&19.2&0\\
			\hline &&&&&&&&&&&&&\\[-2mm]
	1000&1&7&0.14&142&97&0.15&14&997&0.18&0&4999&0.97&0\\
			\hline &&&&&&&&&&&&&\\[-2mm]
	1000&10&7&0.42&159&97&0.48&11&997&0.73&0&4999&5.05&0\\
			\hline &&&&&&&&&&&&&\\[-2mm]
	1000&50&7&2.85&151&97&3.28&12&997&4.46&1&4999&25.4&0\\
			\hline &&&&&&&&&&&&&\\[-2mm]
	5000&1&7&2.24&-&97&2.31&-&997&2.72&-&4999&3.75&-\\
			\hline &&&&&&&&&&&&&\\[-2mm]
	5000&10&7&8.52&-&97&10.2&-&997&12.1&-&4999&20.4&-\\
			\hline &&&&&&&&&&&&&\\[-2mm]
	5000&50&7&154&-&97&161&-&997&274&-&4999&353&-\\
			\hline
		\end{tabular}
\end{center}}
\smallskip
\begin{center}
	{\footnotesize Table 2 \\ Simulation Results for Orthogonal Matrix Generation on HPC Compute Node \par}
\end{center}

{\fontsize{8}{8pt}\selectfont
	\begin{center}
		\begin{tabular}{|c|c||c|c|c||c|c|c||c|c|c||c|c|c|}
			\hline &&&&&&&&&&&&&\\[-2mm]
			$n$&$N$&$p$&$t$&f &$p$&$t$&f &$p$&$t$&f &$p$&$t$&f \\
			\hline &&&&&&&&&&&&&\\[-2mm]
		
			5000&1&7&1.40&146&97&1.54&10&997&1.56&0&4999&1.59&0\\
			\hline &&&&&&&&&&&&&\\[-2mm]
			5000&10&7&4.03&148&97&5.31&13&997&5.37&1&4999&5.91&0\\
			\hline &&&&&&&&&&&&&\\[-2mm]
			5000&50&7&23.4&158&97&26.5&13&997&26.7&1&4999&27.3&0\\
			\hline &&&&&&&&&&&&&\\[-2mm]
			10000&1&7&4.85&156&97&5.12&7&997&5.48&1&4999&5.58&0\\
			\hline &&&&&&&&&&&&&\\[-2mm]
			10000&10&7&15.7&153&97&19.5&12&997&22.1&2&4999&24.7&0\\
			\hline &&&&&&&&&&&&&\\[-2mm]
			10000&50&7&79.6&122&97&87.1&9&997&94.5&1&4999&101&0\\
			\hline &&&&&&&&&&&&&\\[-2mm]
			15000&1&7&11.3&170&97&11.5&13&997&12.3&0&4999&12.6&0\\
			\hline &&&&&&&&&&&&&\\[-2mm]
			15000&10&7&35.2&155&97&43.2&11&997&48.9&0&4999&49.6&0\\
			\hline &&&&&&&&&&&&&\\[-2mm]
			15000&50&7&-&-&97&-&-&997&-&-&4999&-&-\\
			\hline
		\end{tabular}
\end{center}}
\smallskip

Below are some $7\tm 7$ matrices constructed in $Z_{97}$ when the degree $N$ in random generators is $1$. These examples illustrate that the selected coefficients are rather dispersed even for the lowest value of $N$. Observe that these matrices are symmetric and the additional simulations confirm that all matrices $M\in \calO(n,p)$ constructed by the proposed method are symmetric for $N=1$. This indicates that additional regularities can be revealed by further careful exploration of the method.

	{\footnotesize
	$$
	\left[\begin{matrix}
	85&    95&     2     &1    &53    &64    &11 \\
	95    &38    &60    &30    &38    &77    &39\\
	2    &60    &38    &67    &59    &20    &58\\
	1    &30    &67    &83    &78    &10    &29\\
	53    &38    &59    &78    &61    &45    &82\\
	64    &77    &20    &10    &45    &59    &13\\
	11    &39    &58    &29    &82    &13    &29	
	\end{matrix}\right], \left[\begin{matrix}
	37    &76    &22    &96    &25    &71    &73\\
	76    &86    &68    &41    &42    &96    &14\\
	22    &68    &36    &91    &53    &38    &50\\
	96    &41    &91    &63     &2    &60    &33\\
	25    &42    &53     &2    &48    &52    &48\\
	71    &96    &38    &60    &52     &9    &82\\
	73    &14    &50    &33    &48    &82    &17
	
	\end{matrix}\right], \left[\begin{matrix}
	72    &25    &82    &76    &69    &29    &90\\
	25    &18    &48     &9    &12    &43     &3\\
	82    &48    &11    &14    &51    &13    &37\\
	76     &9    &14    &40    &52    &57    &13\\
	69    &12    &51    &52    &38    &76    &82\\
	29    &43    &13    &57    &76    &47    &19\\
	90     &3    &37    &13    &82    &19    &70
	
	\end{matrix}\right]
	$$}

In summary, our study suggests that the proposed method enables the selection of an orthogonal matrix that meets specific constraints through a comprehensive or nearly comprehensive screening process. To substantiate this claim, we provide an example using $4\tm 4$ matrices in $\bZ_5$, where $|\calO(4,5)|$ is known to be $28800$ (see, e.g.,  \cite{McW}). Half of these orthogonal matrices have the determinant $1$, while the other half have the determinant  $-1$. Therefore, it is sufficient to construct $14400$ matrices if we know that all of them have equal determinants.

Because of \eqref{det1N}, if $N$ is the number of the coefficients in randomly generated polynomials $\zeta_i$ in \eqref{Ft}, then the determinant of corresponding ${\bf U}(t)$ is $t^N$ for each $t$ from $\bF$. Therefore, $\det(W)=\det \big({\bf U}(-1)\big)=(-1)^N$ and, for any fixed value of $N$, the method enables the construction of half of the existing orthogonal matrices. For $4\tm 4$ matrices, three random polynomials $\zeta_1, \zeta_2$, and $\zeta_3$ need to be selected, and there are totally $(5^3)^3=1953125$ choices for $N=3$  coefficients (for each $\zeta_i$, $i=1,2,3,$) from $\bZ_5$.  A comprehensive screening through all these cases took around an hour (on the laptop) and $14306$ different matrices from $\calO(4,5)$ were selected. Then we initiated a complete screening (on the HPC) through ($5^4)^3=244140625$ choices for $N=4$, and the process terminated resulting in the selection of all $14400$ different matrices in $5354$ sec $\approx 1.5$ h.
Next, we constructed all ($14400$) different matrices from $\calO(4,5)$ by randomly selecting generators $\zeta_1, \zeta_2, \zeta_3$  with $N=4$ coefficients from $(5^4)^3=244140625$ choices. This random process took $1545$ sec $\approx 0.43$ h and the rate of building these matrices is displayed in Figure 1, indicating that the majority of matrices were constructed in a relatively short period of time. We found that a more efficient strategy was to add all of the left and right even permutations of each newly selected random matrix to the set of already constructed orthogonal matrices. Using this strategy, all $14400$ orthogonal matrices (from the same $244140625$ options of random choices) were constructed in just $8.22$ seconds. The rate of building these matrices is displayed in Figure 2.

\begin{figure}[h]
	\centering
	\includegraphics[width=\textwidth,scale=2]{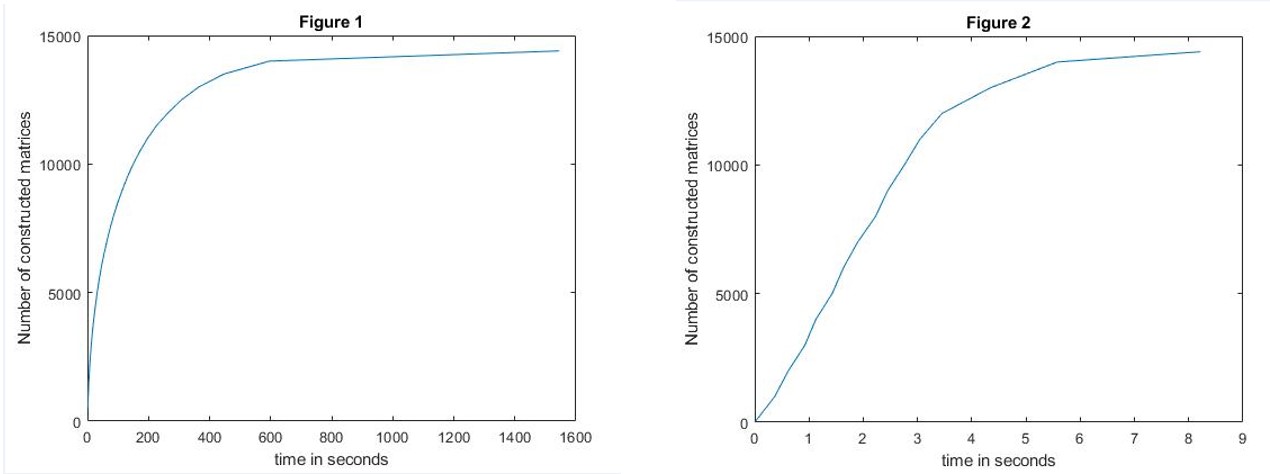} 
\end{figure}

\section{Conclusion}

In this paper, we have introduced a novel method for generating $n\times n$ orthogonal matrices in the Galois field $GF(q)$ that is highly efficient and can handle large values of $n$ and $q$. Our method is based on random selection, but we have shown that it has a low probability of failure for large $q$, as verified through statistical testing. In addition, our method can be used to construct paraunitary filter banks over finite fields. We believe that this approach can be useful in constructing orthogonal matrices that satisfy specific constraints, making it potentially valuable for applications in coding theory and signal processing.

\newpage


\begin{thebibliography}{10}
	
	\bibitem{Cooklev2}
	T.~Cooklev, A.~Nishihara, and M.~Sablatash, \emph{Theory of filter banks over
		finite fields}, Proceedings of APCCAS'94 - 1994 Asia Pacific Conference on
	Circuits and Systems, 1994, pp.~260--265.
	
	\bibitem{TRMI22}
	L.~Ephremidze, A.~Gamkrelidze, and I.~Spitkovsky, \emph{On the spectral
		factorization of singular, noisy, and large matrices by
		{J}anashia-{L}agvilava method}, Trans. A. Razmadze Math. Inst. \textbf{176}
	(2022), no.~3, 361--366.
	
	\bibitem{EL2014}
	L.~Ephremidze and E.~Lagvilava, \emph{On compact wavelet matrices of rank {$m$}
		and of order and degree {$N$}}, J. Fourier Anal. Appl. \textbf{20} (2014),
	no.~2, 401--420.
	
	\bibitem{IEEE2018}
	L.~Ephremidze, F.~Saied, and I.~M. Spitkovsky, \emph{On the algorithmization of
		{J}anashia-{L}agvilava matrix spectral factorization method}, IEEE Trans.
	Inform. Theory \textbf{64} (2018), no.~2, 728--737.
	
	\bibitem{GMJ22}
	L.~Ephremidze and I.~Spitkovsky, \emph{On the generalization of the
		{J}anashia-{L}agvilava method for arbitrary fields}, Georgian Math. J.
	\textbf{29} (2022), no.~3, 353--362.
	
	\bibitem{JMAA22}
	L.~Ephremidze and I.~M. Spitkovsky, \emph{On multivariable matrix spectral
		factorization method}, J. Math. Anal. Appl. \textbf{514} (2022), no.~1, Paper
	No. 126300, 25.
	
	\bibitem{Fekri}
	F.~Fekri, R.~M. Mersereau, and R.~W. Schafer, \emph{Theory of wavelet transform
		over finite fields}, 1999 IEEE International Conference on Acoustics, Speech,
	and Signal Processing. Proceedings. ICASSP99, vol.~3, 1999, pp.~1213--1216.
	
	\bibitem{JL99}
	G.~Janashia and E.~Lagvilava, \emph{A method of approximate factorization of
		positive definite matrix functions}, Studia Math. \textbf{137} (1999), no.~1,
	93--100.
	
	\bibitem{IEEE2011}
	G.~Janashia, E.~Lagvilava, and L.~Ephremidze, \emph{A new method of matrix
		spectral factorization}, IEEE Trans. Inform. Theory \textbf{57} (2011),
	no.~4, 2318--2326.
	
	\bibitem{JLE13}
	\bysame, \emph{Matrix spectral factorization and wavelets}, J. Math. Sci.
	(N.Y.) \textbf{195} (2013), no.~4, 445--454, Translated from Sovrem. Mat.
	Prilozh., Vol. 83, 2012.
	
	\bibitem{Robinson}
	J.~N. MacLaurin and P.~A. Robinson, \emph{Determination of effective brain
		connectivity from activity correlations}, Phys. Rev. E \textbf{99} (2019),
	042404.
	
	\bibitem{McW}
	J.~MacWilliams, \emph{Orthogonal matrices over finite fields}, Amer. Math.
	Monthly \textbf{76} (1969), 152--164.
	
	\bibitem{Vai2}
	S.~M Phoong and P.~P. Vaidyanathan, \emph{Paraunitary filter banks over finite
		fields}, IEEE Transactions on Signal Processing \textbf{45} (1997), no.~6,
	1443--1457.
	
	\bibitem{Poor}
	H.~V. Poor, \emph{Finite-field wavelet transforms}, Information Theory and
	Applications II, Springer Berlin Heidelberg, 1996, pp.~225--238.
	
	\bibitem{RW}
	H.~L. Resnikoff and R.~O. Wells, \emph{Wavelet analysis}, Springer-Verlag, New
	York, 1998, The scalable structure of information.
	
	\bibitem{Vai}
	P.~P. Vaidyanathan, \emph{Multirate systems and filter banks}, Prentice Hall,
	New Jersey, 1993.
	
	\bibitem{Vet1}
	M.~Vetterli, \emph{Filter banks allowing perfect reconstruction}, Signal
	Processing \textbf{10} (1986), no.~3, 219--244.
	
\end{thebibliography}
\providecommand{\bysame}{\leavevmode\hbox to3em{\hrulefill}\thinspace}
\providecommand{\MR}{\relax\ifhmode\unskip\space\fi MR }
\providecommand{\MRhref}[2]{%
	\href{http://www.ams.org/mathscinet-getitem?mr=#1}{#2}
}
\providecommand{\href}[2]{#2}

\end{document}